\documentclass[12pt]{article}
\usepackage{amsfonts,amssymb, amsmath}
\usepackage[ english]{babel}

\def\nn{ \nonumber }
\def\bq{ \begin{equation} }
\def\eq{ \end{equation} }
\def\ben{ \begin{eqnarray} }
\def\en{ \end{eqnarray} }

\def\on#1#2{\mathop{\vbox{\ialign{##\crcr\noalign{\kern2pt}
$\scriptstyle{#2}$\crcr\noalign{\kern2pt\nointerlineskip}
\kern-2pt$\hfil\displaystyle{#1}\hfil$\crcr}}}\limits}

\def\vk{\varkappa}

\begin{document}


\title{Trigonometric Lax matrix for the Kowalevski gyrostat on $so(4)$}
\author{
I.V. Komarov, A.V. Tsiganov\\
\\
\it\small St.Petersburg State University, St.Petersburg, Russia}

\date{}
\maketitle

\begin{abstract}
We present trigonometric Lax matrix and classical $r$-matrix for the
Kowalevski gyrostat on $so(4)$ algebra by using auxiliary matrix
algebras $so(3,2)$ or $sp(4)$.
\end{abstract}

In this note we consider the Kowalevski gyrostat  with the
Hamiltonian
\bq \label{Ham}
H=J_1^2+J_2^2+2J_3^2+2\rho J_3+2y_1,\qquad \rho \in\mathbb R,
\eq
 on a generic orbit of the $so(4)$
Lie algebra with the Poisson brackets
\begin{equation}\label{bundle}
\,\qquad \bigl\{J_i\,,J_j\,\bigr\}=\varepsilon_{ijk}J_k\,, \quad
\bigl\{J_i\,,y_j\,\bigr\}=\varepsilon_{ijk}y_k \,,\quad
\bigl\{y_i\,,y_j\,\bigr\}=\varkappa^2\varepsilon_{ijk}J_k,
\end{equation}
where $\varepsilon_{ijk}$ is the totally skew-symmetric tensor and
$\vk\in \mathbb C$. These brackets are invariant with respect to
transformation $y_i\to ay_i$ and $\vk\to a\vk$, which allows  to
include scaling parameter $a$ into the Hamiltonian.

Because physical quantities $y, J$ in (\ref{Ham}) should be real,
$\vk^2$ must be real too and   algebra (\ref{bundle}) is reduced to
its two real forms  $so(4,\mathbb R)$ or $so(3,1,\mathbb R)$ for
positive and negative $\vk^2$ respectfully and to $e(3)$ for
$\vk=0$.

At $\vk =0$  the Lax matrix  for the Kowalevski gyrostat on $e(3)$
algebra was found in \cite{RS}
\bq\label{lax-e3}
L_0(\lambda)=\left(\begin{array}{rrrrr}0 & J_3 & -J_2 &
\lambda-\dfrac{y_1}{\lambda} & 0 \\ \\ -J_3 & 0 & J_1 &
-\dfrac{y_2}{\lambda} & \lambda \\  \\ J_2 & -J_1 & 0 &
-\dfrac{y_3}{\lambda} & 0 \\  \\ \lambda-\dfrac{y_1}{\lambda} &
-\dfrac{y_2}{\lambda} & -\dfrac{y_3}{\lambda} & 0 & -J_3-\rho  \\
\\0 & \lambda & 0 & J_3+\rho & 0
\end{array}\right)\,.
\eq
In order to construct this matrix  in framework of a general
group-theoretical approach the auxiliary algebra
$\mathfrak g=so(3,2)$ is taken.

At $\vk\neq 0$  the  Lax matrix for the Kowalevski gyrostat on
$so(4)$ is deformation of the  matrix $L_0(\lambda)$ \cite{kst03}
\bq
\label{old-lax} L(\lambda)=Y\cdot L_0(\lambda),\qquad
Y=\mathrm{diag}\left(1,1,1,\dfrac{\lambda^2}{\lambda^2-\vk^2},1\right).
\eq
The corresponding classical $r$-matrix has been constructed in
\cite{ts04}. The algebraic nature of the matrix $L(\lambda)$
(\ref{old-lax}) was mysterious, because diagonal matrix $Y$ does
not belong to the auxiliary $so(3,2)$ algebra.

In this note we present simple trigonometric Lax matrix and the
corresponding trigonometric $r$-matrix  on $so(3,2)$ or $sp(4)$
algebras. Similar to \cite{RS} in order to get the Lax matrices and
the $r$-matrix for the Kowalewski gyrostat on $so(4)$ we  use the
the auxiliary Lie algebra  $\mathfrak g=so(3,2)$ in fundamental
representation.

Let us describe this auxiliary space by using  one antisymmetric
matrix and three symmetric matrices
\[\scriptstyle S_4=\left(\begin{smallmatrix} 0&
0 & 0 & 0 & 0 \\ 0 & 0 & 0 & 0 & 0 \\ 0 & 0 & 0 & 0 & 0 \\ 0 & 0 & 0 & 0 & -1 \\
0 & 0 & 0 & 1 & 0
\end{smallmatrix}\right),\qquad
\scriptstyle Z_1=\left(\begin{smallmatrix} 0 & 0 & 0 & 1 & 0 \\ 0 & 0 & 0 & 0 & 0 \\
0 & 0 & 0 & 0 & 0 \\ 1 & 0 & 0 & 0 & 0 \\ 0 & 0 & 0 & 0 &
0\end{smallmatrix}\right) ,\qquad
\scriptstyle Z_2= \left(\begin{smallmatrix} 0 & 0 & 0 & 0 & 0 \\
 0 & 0 & 0 & 1 & 0 \\  0 & 0 & 0 & 0 & 0 \\  0 & 1 & 0 & 0 & 0 \\ 0
& 0 & 0 & 0 & 0\end{smallmatrix}\right)\,\qquad \scriptstyle
Z_3=\left(
\begin{smallmatrix} 0 & 0 & 0 & 0 & 0 \\ 0 & 0 & 0 & 0 & 0 \\  0 &
0 & 0 & 1 & 0 \\  0 & 0 & 1 & 0 & 0 \\  0 & 0 & 0 & 0 &
0\end{smallmatrix}\right),
\]
which are the generators of the $so(3,2)$ algebra. There are three
other symmetric matrices
\[H_i=[Z_i,S_4]\equiv Z_iS_4-S_4Z_i, \qquad i=1,2,3.\]
and three other antisymmetric matrices
\[S_1=[Z_2,Z_3],\qquad S_2=[Z_3,Z_1],\qquad S_3=[Z_1,Z_2] .\]
Four matrices $S_k$ form maximal compact subalgebra $\mathfrak
f=so(3)\oplus so(2)$ of $so(3,2)$ , whereas six matrices $Z_i$ and
$H_i$ belong to the complimentary subspace $\mathfrak p$ in the
Cartan decomposition $\mathfrak g=\mathfrak f+\mathfrak p$. The
corresponding Cartan involution is given by $\sigma: ~~
X=-X^T$, where $X\in so(3,2)$.

After similarity transformation $L(\lambda)\to
Y^{-1/2}L(\lambda)Y^{1/2}$  of the  Lax $L(\lambda)$
(\ref{old-lax}) and change of the spectral parameter
$\lambda=\vk/\sin\phi$ one gets a trigonometric Lax matrix on the
auxiliary $so(3,2)$ algebra
\bq\label{lax-o4}
L=\dfrac{\varkappa}{\sin\phi}\Bigl(Z_1+\cos\phi
H_2\Bigr)+\sum_{i=1}^3
\left(\cos\phi\,J_i\,S_i-\varkappa^{-1}\sin\phi\,y_i\,Z_i\right)+(J_3+\rho)S_4\,
\eq or \bq\label{lax-55} L=\left(\begin{array}{rrrrr}0 &
\cos\phi\,J_3 & -\cos\phi\,J_2 & \dfrac{\varkappa}{\sin\phi}-
{\dfrac{\sin\phi}{\varkappa}\,y_1}  & 0
 \\ \\-\cos\phi\,J_3 & 0 & \cos\phi\,J_1 & - {\dfrac{\sin\phi}{\varkappa}\,y_2}  &
\dfrac{\varkappa\cos\phi}{\sin\phi} \\ \\ \cos\phi\,J_2 &
-\cos\phi\,J_1 & 0 & - {\dfrac{\sin\phi}{\varkappa}\,y_3}  & 0  \\
\\\dfrac{\varkappa}{\sin\phi}- {\dfrac{\sin\phi}{\varkappa}\,y_1}  & -
{\dfrac{\sin\phi}{\varkappa}\,y_2} & -
{\dfrac{\sin\phi}{\varkappa}\,y_3} & 0 & -J_3-\rho  \\ \\0 &
\dfrac{\varkappa\cos\phi}{\sin\phi} & 0 & J_3+\rho & 0
\end{array}\right).
\eq
In order to consider the real forms $so(4,\mathbb R)$ or
$so(3,1,\mathbb R)$ we have to use trigonometric or hyperbolic
functions for positive and negative $\vk^2$, respectively.

If we put $\phi=\varkappa\lambda^{-1}$ and take the limit $\vk\to 0
$ one gets known rational Lax matrix $L_0(\lambda)$ (\ref{lax-e3})
for the Kowalevski gyrostat on $e(3)$ \cite{RS}
\[
L_0=\lambda(Z_1+H_2)+\sum_{i=1}^3
\left(J_iS_i-\lambda^{-1}\,x_i\,Z_i\right)+(J_3+\rho)S_4\,.
\]

The Lax matrices $L(\phi)$ and $L_0(\lambda)$ are invariant with
respect to the following involutions
$$
L(\phi)\to -L^T(-\phi)\qquad\mbox{\rm and}\qquad L_0(\lambda)\to
-L_0^T(-\lambda),
$$
that are compatible with the Cartan involution $\sigma$. Trigonometric Lax matrix has
also a standard point symmetry $\phi\to \phi+n\pi$, $n\in\mathbb Z$.

It is easy to prove that trigonometric Lax matrix $L(\phi)$
(\ref{lax-o4}) satisfies relation
\bq\label{rpoi}
\left\{\,{\on{L}{1}}(\phi), ~{\on{L}{2}}(\theta)\,\right\}=
\left[r_{12}(\phi,\theta),~{\on{L}{1}}(\phi)\right]
-\left[r_{21}(\phi,\theta),~{\on{L}{2}}(\theta)\,\right]\,.
\eq
with  trigonometric $r$-matrix
\begin{eqnarray}\label{r-trig}
r_{12}(\phi,\theta)&=&\left.
\frac{\sin\phi\sin\theta}{\cos^2\phi-\cos^2\theta}\sum_{i=1}^3\right(
\cos\theta\, H_i\otimes H_i+\cos\phi\, Z_i\otimes
Z_i\\
&-&\left.\frac{\sin\phi\cos\theta}{\sin\theta} S_i\otimes
S_i\right)+ S_3\otimes S_4
+\frac{\cos\phi\sin^2\theta}{{\cos^2\theta-\cos^2\phi}}\,S_4\otimes
S_4\,.\nn
\end{eqnarray}

Here  we use the standard tensor notations
\[{\on{L}{1}}(\phi)=
L(\phi)\otimes 1,\qquad {\on{L}{2}}(\theta)=1\otimes
L(\theta),\qquad
r_{21}(\phi,\theta)=\Pi\,r_{12}(\theta,\phi)\,\Pi\,,
\]
and $\Pi$ is a permutation operator $\Pi X\otimes Y=Y\otimes X\Pi$
of auxiliary spaces. Remark that due to the independent on $\phi$ and $\theta$ item $S_3\otimes S_4$
in (\ref{r-trig}) the inequality  $r_{21}(\phi,\theta)\ne
-r_{12}(\theta,\phi)$ takes place.

If we put $\phi=\varkappa\lambda^{-1}$, $\theta=\varkappa\mu^{-1}$
and take the limit $\vk\to 0$ one gets rational $r$-matrix for the
Kowalevski gyrostat on $e(3)$ algebra \cite{ts04}
\begin{eqnarray*}
\label{}
r_{12}(\lambda,\mu)&=&\dfrac{\lambda\mu}{\lambda^2-\mu^2}\sum_{i=1}^3
\left( H_i\otimes H_i+Z_i\otimes Z_i
-\dfrac{\mu}{\lambda}S_i\otimes
S_i\right)\\
&+&S_3\otimes S_4+\dfrac{\lambda^2}{\lambda^2-\mu^2}S_4\otimes
S_4\,.
\end{eqnarray*}

The well known isomorphism $so(3,2)$ and $sp(4)$ algebras allows
us to consider $4\times 4$ Lax matrix instead of $5\times 5$
matrix (\ref{lax-55}). The four generators $Z_1,Z_2,Z_3$ and $S_4$
may be represented by different $4\times 4$ real or complex
matrices. Below we give one of the possible forms of the $4\times
4$ Lax matrix for the Kowalevski gyrostat
\ben
\label{L4_tr}
L_4(\phi)&=&i\left(%
\begin{array}{cccc}
-e^{-i \phi} J_3& -e^{-i \phi}\vk\sin\phi& \cos\phi ~J_-& 0  \\
\\
e^{-i \phi}\vk\sin\phi & e^{-i \phi} ~J_3&0  &  -\cos\phi ~J_+\\ \\
\cos\phi ~J_+ & 0 &  e^{i \phi} ~J_3& -e^{i \phi}\vk\sin\phi
\\ \\ 0  &  - \cos\phi ~J_- & e^{i \phi} \vk
\sin\phi   &  - e^{i \phi} ~J_3
\end{array}%
\right)+\\
\nn\\
&+&
\vk^{-1}\left(%
\begin{array}{cccc}
0& y_-& 0& y_3 \\ -y_+& 0 & -y_3 & 0\\ 0 &y_3& 0 &-y_+\\
-y_3& 0& y_-& 0
\end{array}%
\right)+
\rho\sin\phi\left(%
\begin{array}{cccc}
-1&0&0& 0\\ 0&1&0&0\\ 0&0&-1&0\\ 0&0&0&1
\end{array}%
\right)\,.\nn
\en
Here $J_\pm=J_1\pm iJ_2$ and $y_\pm=y_1\pm iy_2$.

There are few Lax matrices obtained for different deformations of
known integrable systems from their undeformed counterpart in the
form (\ref{old-lax})  (see \cite{kst03,ts04} and references within).
The main question in construction of these matrices by the  Ansatz
$L=Y\cdot L_0$ (\ref{old-lax}) is a choice of a proper matrix $Y$
for a given rational matrix $L_0(\lambda)$. In this note we show
that this choice is related to transformation of the rational
$r$-matrix to the trigonometric one.

The  trigonometric $r$-matrix (\ref{r-trig}) is constant and the
corresponding  Lax matrix $L(\phi)$ (\ref{lax-o4}) does not
contain ordering problem in  quantum mechanics. Hence  equation (\ref{rpoi}) holds
true in  quantum  case both for Lax matrices
(\ref{lax-o4}) and (\ref{L4_tr}).

\vskip.3cm
\noindent
\textbf{ Acknowledgments}. The authors thank E.K. Sklyanin
 for very useful conversation on the subject of this paper.

 I.V.K. wishes to thank the London Mathematical Society for support his visit to England
 and V.B. Kuznetsov for hospitality at the  University of Leeds.

\end{document}